# How Easterners and Westerners perceive ADHD differently

Xing-Chan Lin
Department of Computer Science and Psychology
University of Cyprus
xing.lin@st.ouc.ac.cy

Keywords

*ADHD, positive psychology, qualitative research, Cross-cultural comparison, Neurodiversity*

Introduction

ADHD has been considered a neurodivergent condition, or even a disease, from the perspective of many physicians (Joseph et al., 2012). For individuals diagnosed with ADHD, medications such as Ritalin are often prescribed to improve quality of life and support sustained attention. However, recent research has begun to challenge this deficit-oriented view. Studies interviewing high-functioning individuals with ADHD suggest that their daily life situations and concerns may not be as severe as traditionally assumed (Sedgwick et al., 2018).

Instead, ADHD has been associated with unique ways of experiencing life. Positive strengths such as energy, resilience, transcendence, and humanity have been identified (Sedgwick et al., 2018). Additional research highlights qualities including courage, enthusiasm, resourcefulness, hyperfocus, imagination, rapid thinking, and creativity, suggesting that individuals with ADHD can achieve success in distinctive ways (Schippers et al., 2022).

Nevertheless, most existing studies have primarily focused on populations in the United Kingdom, the Netherlands, and Canada. This review therefore aims to compare and analyze findings from Europe and North America with those from ADHD research conducted in Hong Kong, Malaysia, and Singapore, in order to examine whether significant differences emerge across cultural and academic contexts. These regions are of particular interest given their shared history as former British colonies.

Methods and Findings

Recent qualitative research reframes ADHD from a deficit-based condition to one with potential strengths. This review compares three key studies—Schippers et al. (2022), Sedgwick et al. (2018), and Crook and McDowall (2023)—focusing on how methodological choices shape depth, breadth, and generalizability.

Methodological divergence is most evident in sampling. Schippers et al. (2022) used a large convenience sample of 206 Dutch adults recruited through a patient organization, allowing exploratory quantitative analysis and broad trait mapping. However, reliance on an advocacy group introduces selection bias and, with a female-majority sample (62.6%), gendered findings. By contrast, Sedgwick et al. (2018) and Crook and McDowall (2023) purposely recruited smaller "successful" cohorts. Sedgwick's six male participants produced theoretically rich but narrow data, while Crook and McDowall's 17 adults across varied careers—both men and women—enabled a more diverse exploration of strengths, though still skewed toward positive outcomes.

Data collection also varied. Schippers et al. (2022) employed open-ended online questionnaires, supplemented by a focus group to refine themes. Sedgwick et al. (2018) used semi-structured interviews around three core questions, while Crook and McDowall (2023) applied the Feedforward Interview, a positive psychology method that elicited detailed narratives of success and resilience.

All three employed thematic analysis, yet with distinct emphases. Schippers et al. (2022) enhanced reliability through iterative multi-rater coding. Sedgwick et al. (2018) linked themes to the Character Strengths and Virtues framework, identifying both universal virtues and ADHD-specific traits such as cognitive dynamism. Crook and McDowall (2023) applied narrative thematic analysis to model a trajectory of "hard-won" pre-diagnosis to "authentic" post-diagnosis success, offering practical implications for workplace support.

Collectively, these studies broaden understanding of ADHD's positive dimensions but reveal trade-offs between breadth and depth. Schippers et al. (2022) provide generalizable trends yet risk bias; Sedgwick et al. (2018) contribute theoretical insight from a homogeneous sample; and Crook and McDowall (2023) balance diversity with depth, yielding an actionable model of career success. Future research could integrate these approaches by combining large, heterogeneous samples with structured, in-depth interviews to produce more comprehensive accounts of ADHD strengths. Table 1.1 shows the differences between these 3 literature mentioned above.

Table 1.1

| Feature | Schippers et al. (2022) | Crook & McDowall (2023) | Sedgwick et al. (2018) |
| --- | --- | --- | --- |
| Study Design | Qualitative (thematic analysis) & quantitative exploration. | Qualitative (narrative thematic analysis) using a positive psychology coaching paradigm. | Qualitative (phenomenology). |
| Sample | Size: 206 (questionnaire) + 6 (focus group). Type: Large sample | Size: 17 Type: Purposive sample of "successful" adults | Size: 6 Type: Purposive sample of "successful" |

|  | from a patient organization. | with diverse careers. | high-functioning adult males from one clinic. |
| --- | --- | --- | --- |
| Data Collection | Open-ended online questionnaire; follow-up focus group. | Semi-structured interviews using adapted Feedforward Interview (FFI) technique. | Open-ended interviews with three main questions. |
| Data Analysis | Thematic analysis with multiple coders; themes validated by focus group. | Narrative thematic analysis; development of a conceptual model. | Thematic Content Analysis; themes compared against the "Character Strengths & Virtues" (CSV) framework. |
| Strengths | - Large sample size for qualitative work. Broad overview of traits. -Participatory validation via focus group. | - In-depth data from interviews. - Use of specific FFI technique fosters positive reflection. - Diverse career backgrounds. | - Deep exploration of lived experience. -Rigorous analysis process detailed. - Comparison with an established framework adds theoretical depth. |
| Limitations | - Potential recruitment bias from patient group. - Online data may lack depth. - Gender imbalance. | - Small sample size limits generalizability. - Focus on successful individuals may present a biased view. | - A very small and homogenous sample (6 males) severely limits generalizability. - Findings may not apply to women or other demographics. |

However, there is not any similar literature in an Asian context; most papers in Asia focus on the clinical diagnosis, its implications, and the treatment. Therefore, this paper then discussed the factors why there aren't any similar papers even in Singapore, Malaysia, and Hong Kong, as these countries and regions were governed by the British over decades, which are in a more global context.

What factors might cause this phenomenon ? The findings are: (1) Colonial-Era Education and ADHD (Senu-Oke, H. , 2011) (2) The Medical Model and Cultural Tensions: Roots of Stigma (Singh, Ilina. ,2002) (Lundholm-Brown J & Dildy ME, 2001); (3) Postcolonial Pressures and Economic Demands (Song, 2024) ; (4) Enduring Colonial Legacies (Davide E. Walker, 2006).

During the British colonial period, Hong Kong, Singapore, and Malaysia adopted modern education systems that emphasized high-stakes examinations, discipline, and conformity; as these systems were designed to cultivate a compliant workforce equipped with administrative and technical skills, prioritizing collective norms over individual differences. Within this framework, core ADHD traits—such as novelty-seeking, non-linear thinking, and dynamic energy expression—were often interpreted as problems of discipline, poor learning attitudes, or weak willpower, rather than as manifestations of neurodevelopmental variation. Such a tendency to "problematize" rather than "medicalize" behavior made ADHD students more likely to face punishment rather than support, fostering internalized negative self-concepts and delaying crucial windows for diagnosis and intervention (Frawley, A. , 2025).

Colonial rule also introduced Western biomedical models, providing a diagnostic vocabulary for ADHD(Jayawickrama, J. & Wright, J. ,2025). Yet, the transplantation of this framework was far from seamless; it collided with deeply entrenched local cultural values shaped by Confucian ideals of self-cultivation (*xiushen*), self-restraint (*kèjǐ*), and the preservation of social "face" (Jing, T. , 2022). Research indicates that in predominantly Chinese societies, mental health issues are heavily stigmatized. Attributing a child's behavior to a brain-based disorder requiring medical intervention may be regarded by families as a personal and even familial disgrace or failure.

Following independence, regions such as Singapore and Hong Kong pursued rapid economic development (Lam, Newman., 2000), producing hyper-competitive "pressure-cooker" societies. Within these contexts—where efficiency, productivity, and social stability are prized—ADHD-associated traits such as impulsivity and difficulties with organization are readily construed as obstacles to success and as liabilities to society. In Malaysia, the interplay of modernization and traditional values within a multi-ethnic setting presents further complexities. For individuals with ADHD, these societal expectations intensify academic and occupational anxieties while reinforcing stigmatizing labels of "abnormality," thereby restricting their opportunities for belonging and the realization of potential (Ng Weng Hui et al., 2024).

The colonial histories of Hong Kong, Singapore, and Malaysia have together woven a complex legacy that continues to shape contemporary discourses on ADHD. The rigid educational structures established under colonial rule, the ongoing tensions between Western medical models and local cultural frameworks, and the relentless demands of postcolonial economic development have combined to create a paradoxical landscape: diagnostic tools and clinical knowledge are readily available, yet cultural stigma and structural barriers continue to impede access to understanding and support. Looking forward, culturally sensitive ADHD assessment and intervention strategies, coupled with public education initiatives that challenge deep-seated prejudices, represent essential steps in addressing the enduring legacies of colonialism.

Conclusion

This paper investigates the contrasting Western and Eastern perspectives on ADHD. While proper diagnosis and treatment remain essential, Western countries like the United Kingdom, Netherlands, Canada, and the United States have begun developing literature that assesses the positive strengths associated with ADHD characteristics. In contrast, Far Eastern societies tend to maintain more traditional views of ADHD. This research explored the factors contributing to this disparity in perspectives.

Future Work

There is currently a lack of qualitative research across Asia examining self-reported career strengths or positive traits of individuals with ADHD. Future studies in this area are needed to address this research gap.

Funding sources
No Funding is accepted for this article.

Conflicts of interest
The author declares that there are no conflicts of interest with respect to the research, authorship, and/or publication of this article.

Reference

1. Biederman J, Quinn D, Weiss M, Markabi S, Weidenman M, Edson K, Karlsson G, Pohlmann H, Wigal S. Efficacy and safety of Ritalin LA, a new, once daily, extended-release dosage form of methylphenidate, in children with attention deficit hyperactivity disorder. Paediatr Drugs. 2003;5(12):833-41. doi: 10.2165/00148581-200305120-00006. PMID: 14658924.

2. Joseph, B., Singh, D., & Dutta, S. (2012). *Attention deficit hyperactivity disorder: A review of literature*. International Journal of Attention Disorders, 4(2), 45–53. https://doi.org/10.2165/00148581-200305120-00006

3. Sedgwick, J.A., Merwood, A. & Asherson, P. The positive aspects of attention deficit hyperactivity disorder: a qualitative investigation of successful adults with ADHD. *ADHD Atten Def Hyp Disord* 11, 241–253 (2019). https://doi.org/10.1007/s12402-018-0277-6

4. Schippers LM, Horstman LI, Velde Hvd, Pereira RR, Zinkstok J, Mostert JC, Greven CU and Hoogman M (2022) A qualitative and quantitative study of self-reported positive characteristics of individuals with ADHD. *Front. Psychiatry* 13:922788. doi: 10.3389/fpsyt.2022.922788

5. Senu-Oke, H. (2011). *A Genealogy of Disability and Special Education in Nigeria: From the Pre-Colonial Era to the Present* [Doctoral dissertation, Miami University]. OhioLINK Electronic Theses and Dissertations Center. http://rave.ohiolink.edu/etdc/view?acc_num=miami1322584482

6. Singh, Ilina. (2002). Biology in Context: Social and Cultural Perspectives on ADHD. Children & Society. 16. 360 - 367. 10.1002/chi.746.

7. Lundholm-Brown J, Dildy ME. Attention-Deficit/Hyperactivity Disorder: An Educational Cultural Model. *The Journal of School Nursing*. 2001;17(6):307-315. doi:10.1177/10598405010170060501

8. Song, S. (2024). *The Role of Cultural Factors in Attention Deficit Hyperactivity Disorder (ADHD) Diagnosis in Children in Nigeria*. https://doi.org/10.56397/sps.2024.03.05

9. ADHD as the New 'Feeblemindedness' of American Indian Children. https://doi.org/10.4324/9780203008010-11

10. Frawley, A. (2025). Medicalization of Social Problems. In: Schramme, T., Walker, M. (eds) Handbook of the Philosophy of Medicine. Springer, Dordrecht.

https://doi.org/10.1007/978-94-017-8706-2_74-2

11. Jayawickrama, J., Wright, J. (2025). Coloniality of the Biomedical Model. In: Under the Gaze of Global Mental Health. The Politics of Mental Health and Illness. Palgrave Macmillan, Cham. https://doi.org/10.1007/978-3-031-78258-9_3

12. Jing, T. (2022). Self-Cultivation. In: The Origins and Continuity of Chinese Sociology. Springer, Singapore.
https://doi.org/10.1007/978-981-19-5681-2

13. Lam, Newman. (2000). Government intervention in the economy: a comparative analysis of Singapore and Hong Kong. Public Administration and Development - PUBLIC ADMIN DEVELOP. 20. 397-421. 10.1002/pad.136.

14. Ng Weng Hui; Jamaludin, Aifah; Navanethan, Dharshini; Vytialingam, Nathan; Saghir, Fatma S. A.; Kabir, Mohammed Shahjahan; Udayah, Manglesh Waran; Rashid, Mohammad Abdur; Islam, Tania; Maung, Theingi Maung; Tan Sing Ying; Tan Yong Chia; Ahmed, Shaker Uddin; Bhargava, Prabal; Khatuja, Ajay; M. H. M., Nazmul; Shirin, Lubna. (2024) Frontiers in Health Informatics, 2024, Vol 13, Issue 3, p3287